\newcounter{myctr}

\documentclass{ws-acs}
\usepackage{url}
\begin{document}

\makeatletter
\def\@biblabel#1{[#1]}
\makeatother

\markboth{P. Pohorecki, J. Sienkiewicz, M. Mitrovi\'{c}, G. Paltoglou, J. A. Ho{\l}yst}{Statistical Analysis of Emotions and Opinions at Digg Website}

%
\catchline{}{}{}{}{}
%

\title{STATISTICAL ANALYSIS OF EMOTIONS AND OPINIONS AT DIGG WEBSITE}

\author{PIOTR POHORECKI}
\address{Faculty of Physics, Centre of Excellence for Complex Systems Research, Warsaw University of Technology, Koszykowa 75, 00-662 Warszawa, Poland\\
piotr.pohorecki@gmail.com}

\author{JULIAN SIENKIEWICZ}
\address{Faculty of Physics, Centre of Excellence for Complex Systems Research, Warsaw University of Technology, Koszykowa 75, 00-662 Warszawa, Poland\\
julas@if.pw.edu.pl}

\author{MARIJA MITROVI\'{C}}
\address{Department of Theoretical Physics, Jo\u{z}ef Stefan Institute, P.O. Box 3000, 1001 Ljubljana, Slovenia\\
marija.mitrovic@ijs.si}

\author{GEORGIOS PALTOGLOU}
\address{School of Technology, University of Wolverhampton, Wulfruna Street, Wolverhampton
WV1 1LY, United Kingdom\\
g.paltoglou@wlv.ac.uk }

\author{JANUSZ A. HO{\L}YST}
\address{Faculty of Physics, Centre of Excellence for Complex Systems Research, Warsaw University of Technology, Koszykowa 75, 00-662 Warszawa, Poland\\
jholyst@if.pw.edu.pl}

\maketitle

\begin{history}
\received{(received date)}
\revised{(revised date)}
\end{history}

\begin{abstract}
We performed statistical analysis on data from the Digg.com website, which enables its users to express their opinion on news stories by taking part in forum-like discussions as well as directly evaluate previous posts and stories by assigning so called "diggs". Owing to fact that the content of each post has been annotated with its emotional value, apart from the strictly structural properties, the study also includes an analysis of the average emotional response of the posts commenting the main story. While analysing correlations at the story level, an interesting relationship between the number of diggs and the number of comments received by a story was found. The correlation between the two quantities is high for data where small threads dominate and consistently decreases for longer threads. However, while the correlation of the number of diggs and the average emotional response tends to grow for longer threads, correlations between numbers of comments and the average emotional response are almost zero. We also show that the initial set of comments given to a story has a substantial impact on the further "life" of the discussion: high negative average emotions in the first 10 comments lead to longer threads while the opposite situation results in shorter discussions. We also suggest presence of two different mechanisms governing the evolution of the discussion and, consequently, its length.
\end{abstract}


\section{Introduction}
Although the concept of physical modelling of social processes is older than the idea of statistical modelling of physical phenomena (dating back to the times of Laplace, Comte or Stuart Mill) \cite{ball}, it has not been until recent years that the methods and models of physics became widely (and successfully) employed to the description of social phenomena. Thanks to its very general conceptual framework, the field of statistical physics has proven a tool of exceptional use in this regard \cite{cast}.

The lively increase observed in the last fifteen years in the number of papers and works concerning the field in question was due to many factors. Firstly, with the advances of computational powers and storage capabilities large digital databases have become available to scientists. Secondly, owing to the rapid development of the Internet unprecedented social processes appeared and prompted more and more physicists to turn their attention to the rising domain of sociophysics.

Exploring the behaviour of complex systems comprising humans rather than particles, physicists have tackled so far a number of phenomena: from spontaneous formation of a common culture and its dissemination through evolution of opinions and crowd behaviour to language dynamics \cite{cast,axel,sznajd}. Very recently studying and modelling of collective emotions in Internet communities also gained considerable interest.

Collective emotions are relatively straightforward to notice in online communities. Highly emotional discussions are usually connected with very exciting, controversial or tragic events. When many users take part in such highly emotional communication we call this phenomenon collective emotion. The studies of collective emotions in online communities comprises two major areas. First is \textit{sentiment analysis} - computational methods to extract emotional content of a written text and to classify this content according to a set of possible dimensions. Second - building mathematical models of the emergence of collective emotions based on the psychological and sociological body of knowledge on emotion. Computational simulations of the models and the comparison of their results to empirical data verifies the validity of these models.

During the last two years several papers were published presenting studies about the presence of collective emotions in the Internet - both empirical \cite{bosa1,frank,bosa2,bosa3,ania2,ania3,weronski,frank2} as well as theoretical \cite{aga,ania1,rank} let alone the ones that stress the applicability of the results \cite{app1,app2,app3}. The objective of this very paper was to analyse available Digg.com dataset trying to find regularities and relations that would give insight into the dependencies between the emotional content of online discussions and the opinions issued by its users.

\section{Data and methods}
Internet discussion participants do not transmit their emotions directly, they communicate via text messages which, depending on their emotional content, may induce certain emotional responses in readers. Information on emotions of individuals can usually be inferred from the physiological signals sent by their body. However, in online communities this is not the case; we are left only with textual statements from discussion participants. The question arises how to infer from these statements information regarding emotions.

Before trying to detect emotions, one needs to decide on how to measure them. A well-established psychological theory of emotion, the circumplex model, is commonly used in sociophysics modelling. It takes into account 2 dimensions of emotion: \textit{valence} and \textit{arousal}. The former indicates how positive or negative the emotion is, the latter - the level of personal activity caused by that emotion (from lethargic to hyperactive) \cite{frank}. Depending on different methods \textit{sentiment analysis} can usually provide \textit{valence} or a specific combination of the two values \cite{mike1}. Different procedures of emotion detection are in use. The analysis in \cite{mike1} and \cite{sobkowicz} engaged manual (human) annotation, but this method allows only for a limited number of textual statements to be assessed. Much larger amount of data can be processed with the use of automatic annotation developed within the already mentioned field of sentiment analysis.

Application of sentiment analysis (also known as \textit{opinion mining}) to the detection and classification of emotions is a development of the field which initially concentrated on extracting opinions \cite{mike1}. In the last ten years this area of research has seen a substantial growth \cite{pang} gaining a lot of attention from industry and academia alike. This was due to the phenomena of Web 2.0 leading to an unprecedented increase in the amount of online content generated by regular users, rather than website owners or publishers. The information contained in user-generated content (UGC) could be of pivotal importance to firms and institutions. Hence, first efforts of sentiment analysis focused on analysing multiple movie reviews or comments regarding manufacturer's products with the purpose to determine which features receive most positive and negative feedback. The two fundamental tasks of 
sentiment analysis are: (i) identifying whether a text is objective or subjective (i.e. contains facts or opinions/emotions) (ii) determining its subjective polarity (i.e. identifying how positive/negative it is).

In case of this study we used the following approach: during the training phase, the program is fed with a set of documents classified by humans for emotional content (positive, negative or objective) from which it learns the characteristics of each type. Afterwards, during the second phase, the algorithm applies obtained sentiment classification knowledge to previously unseen documents. We trained a hierarchical Language Model \cite{hLM,bosa2} on the Blogs06 collection \cite{blogs} and applied the trained model to the Digg data. Each post is initially classified as objective or subjective and in the latter case, it is further classified in terms of its polarity, i.e., positive or negative. Each level of classification applies a binary Language Model \cite{lm1,bosa2}. Eventually posts are annotated with a single value $e =-1, 0$ or $1$ to indicate their valence (negative, neutral or positive, respectively). The dataset was obtained by a complete crawl of the Digg site \cite{digg} for months February, March and April 2009. Data concerning stories submitted during that period was collected. Information on users, diggs and comments relevant to the stories was also gathered. The most fundamental properties of the dataset are shown in Table \ref{tab_fund}.

\begin{table}[!hh]
\tbl{Fundamental statistics of the dataset\label{tab_fund}}
{\begin{tabular}{@{}ccccc@{}} 
\toprule
Stories/posts & Comments & Users & Stories per user & Comments per user\\ 
\colrule
1195808 & 1646153 & 484985 & 2.47 & 3.39\\
\botrule
\end{tabular}}
\end{table}

An example of an online user-generated content (USG) rating network, Digg.com relies on users to submit and moderate news stories. Each newly-submitted story goes to the \textit{Upcoming} section, which is the place where users browse and vote for (or using website's nomenclature: \textit{digg}) the stories they like most. Once the story fulfils special criteria it gets promoted and is moved to the \textit{Popular} section displayed as website's \textit{front page}. The exact promotion algorithm is not known to the public (and changes on regular basis), but the number of votes (diggs) and the rate at which a story receives them are the most important factors \cite{lerman1,zhu,szabo,lerman2,rangwala}. The order in which promoted stories are featured at the \textit{front page} is also a subject to the algorithm. The most interesting and relevant stories (according to the promotion mechanism) are placed at the top of the page.

\begin{figure}[!hh]
\centerline{\epsfig{file=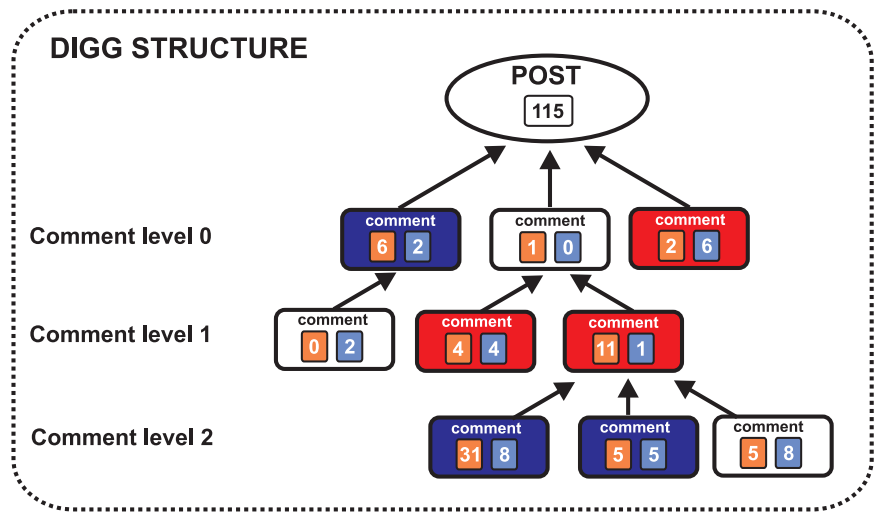,width=\textwidth}}
\caption{Digg structure as present in the gathered data. Each story starts with a post (an empty ellipse) that can obtain diggs (a box inside the ellipse). The post can be commented as well as comments can, however no deeper than to the second level. The comments (rectangular boxes) themselves also obtain diggs but contrary to the post they can be both positive (diggs up) or negative (diggs down). Each comment is subject to emotional classification, shown with different colours of the comment box.}\label{fig:struct}
\end{figure}

Apart from receiving diggs, each story can be commented. Comments, in turn, can obtain diggs (approvals), but in addition they can also be disapproved of by getting \textit{thumbs down/diggs down} (the activity called \textit{buring}). In the context of analysing the emotional content of websites, Digg's voting system seems of pivotal importance. In addition to emotional classification carried out by Wolverhampton partners, the dataset allows for analyses of the number of \textit{diggs up}/\textit{diggs down} which to some extent reflects how interesting and/or emotionally engaging the assessed story/comment was. A schematic plot of the Digg structure is shown in Fig. \ref{fig:struct}.

\section{Structural properties}

\subsection{Threads' lifetime}
\begin{figure}[!ht]
\centerline{\epsfig{file=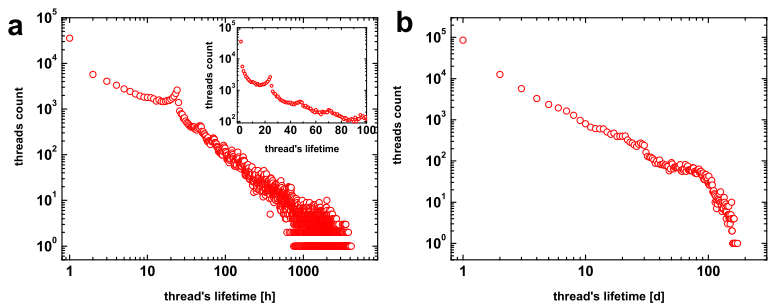,width=0.8\textwidth}}
\caption{(a) Log-log plot of the histogram of threads' lifetime given in hours. The inset shows the same data in a semi-logarithmic scale. (b) Log-log plot of the histogram of threads' lifetime given in days. Both hours and days are rounded up to integer values.}
\label{fig:tt_dist}
\end{figure}

We define thread's lifetime as the time that elapsed between the first and the last comment in a thread. Using comments' timestamps (exact date and time) threads' lifetimes were calculated. Their histograms (using different ranges and time scales) are plotted in Fig. \ref{fig:tt_dist}a. They reveal clear increase of threads counts for lifetimes of 24 hours and mulitples of this period (inset in Fig. \ref{fig:tt_dist}a). Also, using different time scale, the character of the graph seems to change around the value of 30 days (Fig. \ref{fig:tt_dist}b).

Most certainly this behaviour is due (at least in part) to the promotion mechanism of the site and the presentation of top ranked material. Next to thematic categories, Digg.com allows for viewing the most popular stories within a specific period of time. At the front page an Internet user may choose to browse through most recent stories, top in 24 hours, 7 days, 30 days or 365 days.
\subsection{Comments distributions}
\begin{figure}[!ht]
\centerline{\epsfig{file=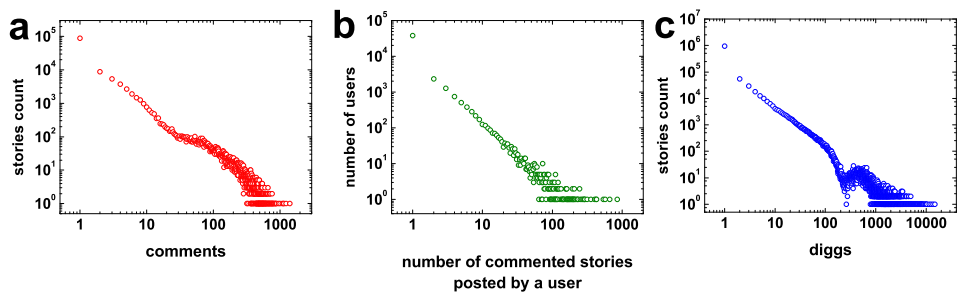,width=\textwidth}}
\caption{(a) Log-log histogram of the number of comments for all data. (b) Log-log histogram of the users' productivity i.e. the number of users that posted a certain amount of stories in {\it Digg.com}. Only stories that acquired at least one comment were taken into account. (c) Log-log plot of number of stories that obtained a certain amount of diggs.}
\label{fig:distr}
\end{figure}

Histogram of the number of comments for all data (Fig. \ref{fig:distr}a) show two distinct distributions: for lower values a power law can be observed, and then starting around 20th comment a significantly different distribution takes over. One can hypothesise that the two distributions might be generated by two distinct classes of users (e.g. regular ones and spammers or advertising/marketing professionals). In order to verify that statement graphs presenting users' productivity were plotted (Fig. \ref{fig:tt_dist}b). Productivity was measured in the number of commented stories (stories that received at least 1 comment) and it follows a single power law relation indicating meaningful presence of only one class of users. Otherwise we would see multiply power law (or other) distributions, one for each group.
\subsection{Diggs distributions}
Similarly to comments, diggs histogram - presented in Fig. \ref{fig:distr}c displays two distinct distribution. Plots first start with power law and then evolve into Gaussian peak. Power law relation may be explained by a preferential process \cite{krap} - a phenomenon quite common in complex systems, responsible for fat-tailed distributions including power-laws \cite{ba,kr}. Obtained histograms represent real data and they are not an outcome of a simulation, however the same mechanism of preferential attachment was at work here. The more diggs the main post (story) obtained in a specific period of time the better promoted it was through the Digg.com algorithm (front-page placement, higher ranking position etc).

\begin{table}[!hh]
\tbl{Number of users engaged in different activities.\label{tab_users}}
{\begin{tabular}{@{}lr@{}}
\toprule
activity & number of users\\ 
\colrule
gave at least 1 digg & 460584\\
\colrule
posted at least 1 comment & 137183\\
submitted at least 1 story & 218686\\
posted at least 2 comments & 77804\\
submitted at least 2 stories & 108068\\
\colrule
submitted at least 1 story, each of the stories had at least 1 comment & 44962\\
\colrule
submitted at least 1 story, each of the stories had at least 4 comment & 7128\\
submitted at least 1 story, each of the stories had at least 8 comment & 3515\\
submitted at least 1 story, each of the stories had at least 15 comment & 2224\\	
\botrule
\end{tabular}}
\end{table}

\section{Average emotional response}
Different posted materials trigger different emotional responses. In order to measure overall reaction we introduce a new quantity. We define average emotional response to a main post (story) as a mean value of the emotional content of the comments (all levels) submitted to this story
\begin{equation}
\langle e \rangle = \frac{1}{N} \sum_{i = 1}^{N} e_i 
\end{equation}
where $e_i \in \{-1,0,1 \}$ is the emotional content of the $i$-th comment and $N$ is the number of comments (all levels) submitted to a given story.

To determine responses to materials submitted by individual users (or the ones published at individual websites), we group together threads started by the same user (or originating from the same website) and calculate averages of $\langle e \rangle$ in those groups. In the following three subsections average response to individual posts (that is $\langle e \rangle_{thread}$ in each thread) as well as to individual users $\langle e \rangle_{user}$ and websites $\langle e \rangle_{website}$ will be presented.

\begin{figure}[!ht]
\centerline{\epsfig{file=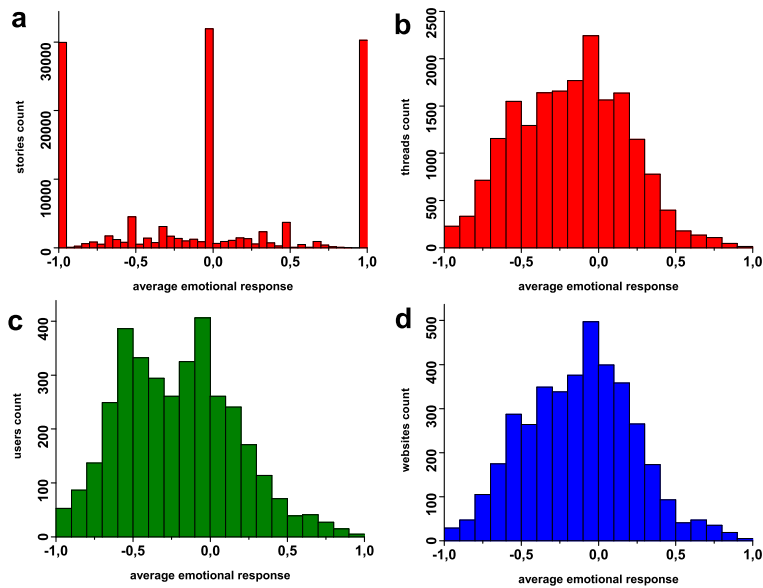,width=0.8\textwidth}}
\caption{Histograms of the average emotional response. (a) Average emotional response of threads $\langle e \rangle_{thread}$, number of bins $N_{bins}=50$, all data. (b) Average emotional response of threads $\langle e \rangle_{thread}$, number of bins $N_{bins}=20$, threads with a threshold $N_{comments} \geq 8$. (c) Average emotional response to material submitted by specific users $\langle e \rangle_{user}$, number of bins $N_{bins}=20$, threads with a threshold $N_{comments} \geq 8$. (d) Average emotional response to a content from specific websites $\langle e \rangle_{website}$, number of bins $N_{bins}=20$, threads with a threshold $N_{comments} \geq 8$.}
\label{fig:avr_dist}
\end{figure}

\subsection{Threads}
Analysis concerned only commented stories, i.e. those which initiated a thread. There were 129 998 such main posts (stories). As can be seen from Fig. \ref{fig:avr_dist}a, due to a large number of very short threads in the whole data there is a prevalence of $\langle e \rangle$ having exact (it employs 1000 bins) value of either -1,0 or 1. Similarly, smaller peaks correspond to values of $\pm\frac{1}{2}$,$\pm\frac{1}{3}$,$\pm\frac{1}{4}$,$\pm\frac{3}{4}$, etc. which are the only possible results for short (and quite numerous) threads. When we take into account a subset comprising threads of the length of 8 or more comments (Fig. \ref{fig:avr_dist}b), we eliminate peaks for the three values and get a distribution resembling normal distribution shifted slightly to negative values.

\subsection{Users}
Collected data enabled to track material submitted by specific users and to measure average overall emotional response to their posts. During the crawl there were 484 985 registered and active users who either: posted a story, wrote a comment or dug some material. Table \ref{tab_users} shows the number of users who engaged in different activities.

It can be seen from the table that more users were interested in submitting a story than posting a comment. This is probably due to fact that the primary appeal of Digg.com to vast majority (so-called 'light users') is to share content of interest with others, the desire to exchange comments at the website being of slightly lower importance. 

In order to cut off peaks for -1,0 and 1, thresholds in the number of comments for a given story were used. The number of stories posted by a user did not seem a good threshold as there were users who posted many stories of which few were commented. For example: user with ID 59919, who posted links to a web site on golf in the UK submitted 1402 stories (the greatest number of all users) of which only 5 were commented with the total of 6 comments, i.e. 4 stories received 1 comment and 1 story - 2 comments. He posted none own comments.

Figure \ref{fig:avr_dist}c shows that histograms for users are similar in character: without a threshold in the number of comments there are 3 distinct peaks for -1,0 and 1. Once the threshold is introduced, we can distinguish two peaks for values near 0 and -0.5.

\subsection{Websites}
As was the case with individual users, collected data enabled also to measure emotional response to content from specific websites (e.g. {\it www.youtube.com}, {\it www.nytimes.com}, {\it news.bbc.co.uk}). Top 50 most popular websites (in terms of the number of times their content appeared at {\it Digg.com}) are presented in Table \ref{tab:websites}. The outright winner is video-sharing website {\it www.youtube.com}, followed by {\it www.examiner.com} (citizen journalism) and a number of professional online newspapers/news outlets ({\it www.nytimes.com}, {\it news.bbc.co.uk}, {\it www.cnn.com}, {\it www.telegraph.co.uk}) and news aggregators ({\it www.huffingtonpost.com}, {\it news.yahoo.com}). It is worth noting that thanks to taking into account only commented posts we eliminate the impact of material submitted for advertising or marketing purposes.
 
\begin{table}[!hh]
\tbl{Top 50 most popular websites (in terms of the number of appearance).\label{tab:websites}}
{\begin{tabular}{@{}clrrc@{}}
\toprule
Rank & Website & Counts  & Comments & $\langle e \rangle_{website}$ \\
\colrule
1&www.youtube.com &6808&76433&0.067\\
2&www.examiner.com &2512&14280&-0.131\\
3&www.nytimes.com &1465&34719&-0.280\\
4&www.huffingtonpost.com &1451&35478&-0.338\\
5&news.bbc.co.uk &1222&15092&-0.301\\
6&news.yahoo.com &1066&21225&-0.394\\
7&www.cnn.com &965&17773&-0.373\\
8&www.telegraph.co.uk &884&32447&-0.232\\
9&www.reuters.com &838&12928&-0.342\\
10&rawstory.com &729&14136&-0.563\\
11&www.washingtonpost.com &688&12435&-0.410\\
12&www.flickr.com &666&20378&0.091\\
13&www.foxnews.com &661&6675&-0.451\\
14&arstechnica.com &658&16697&0.064\\
15&www.squidoo.com &632&825&0.310\\
16&online.wsj.com &558&13298&-0.300\\
17&i.gizmodo.com &549&15481&0.132\\
18&www.time.com &532&22557&-0.243\\
19&www.dailymail.co.uk &505&16299&-0.245\\
20&www.alternet.org &497&6446&-0.389\\
21&www.msnbc.msn.com &479&16409&-0.239\\
22&www.guardian.co.uk &475&12341&-0.311\\
23&news.cnet.com &468&13633&0.064\\
24&www.worldnetdaily.com &459&4214&-0.483\\
25&blog.wired.com &415&11508&0.009\\
26&www.chicagotribune.com &402&9077&-0.121\\
27&www.latimes.com &379&8885&-0.313\\
28&www.collegehumor.com &371&9034&0.138\\
29&www.opednews.com &348&644&-0.249\\
30&www.breitbart.com &338&6899&-0.421\\
31&www.politico.com &325&8571&-0.499\\
32&www.news.com.au &317&10085&-0.127\\
33&bnp.org.uk &314&1753&-0.556\\
34&hubpages.com &301&439&0.189\\
35&www.break.com &295&4068&0.026\\
41&www.engadget.com &257&10362&0.186\\
42&www.usatoday.com &247&4658&-0.262\\
43&www.dailykos.com &243&4043&-0.384\\
44&www.slate.com &240&3006&-0.157\\
45&www.google.com &240&2920&-0.246\\
46&www.ebaumsworld.com &239&578&-0.003\\
47&www.salon.com &237&6367&-0.459\\
48&abcnews.go.com &234&6170&-0.257\\
49&blog.propertynice.com &226&248&0.338\\
50&www.theonion.com &221&4043&0.118\\
\botrule
\end{tabular}}
\end{table}

Out of the top 50 most popular websites those with the lowest and highest value of average emotional response were listed in Table \ref{tab:neg} and Table \ref{tab:pos} respectively. As can be seen, websites dealing with politics generate most negative emotions (blogs and news websites with political bias, website of the British National Party) while those concentrated on gadgets and technology/giving advice/with humorous content receive most positive responses. Similar conclusions were arrived at in a paper by P. Sobkowicz and A. Sobkowicz \cite{sobkowicz}. The authors analysed data from the \textit{Politics} section of discussion fora at one of the most popular Internet portals in Poland - {\it www.gazeta.pl}. Using human assessment of comments for a sample of discussion threads they noted that aggressive comments (disagreeing, provocative or invectives) accounted for 75\% of communication between \textit{Politics} forum users. This was not the case in \textit{Sports} or \textit{Science} forum. Moods and opinions of \textit{Sports} section participants turned out to be usually similar while discussions on science, if happened to be longer, tended to be more factual in character.

\begin{figure}[!ht]
\centerline{\epsfig{file=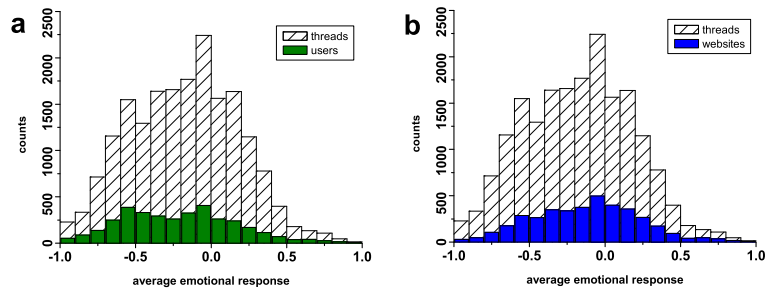,width=0.8\textwidth}}
\caption{Comparison of average emotional histograms. (a) threads and users, (b) threads and websites. Number of bins $N_{bins}=20$, all threads with a threshold $N_{comments} \geq 8$.}
\label{fig:avr_bg}
\end{figure}

\begin{table}[!hh]
\tbl{Top 15 websites (out of top 50 most popular) with the lowest $\langle e \rangle_{website}$.\label{tab:neg}}
{\begin{tabular}{@{}clccp{5cm}@{}}
\toprule
Rank & Website & Counts rank & $\langle e \rangle_{website}$ & Description\\
\colrule
1&rawstory.com &10&-0.563&liberal news, politics and blogs\\
2&bnp.org.uk &33&-0.556 & British National Party\\
3&www.politico.com &31&-0.499 & American political journalism, conservative and Republican bias\\
4&www.worldnetdaily.com &24&-0.483& news and associated content, American conservative perspective\\
5&www.salon.com &47&-0.459&online magazine, focuses on U.S. politics, criticized for its left-leaning content\\
6&www.foxnews.com &13&-0.451& news channel persived as promoting conservative political positions\\
7&www.breitbart.com &30&-0.421&news site, its Blog \& "Network" links tend to run to the right within the U.S. political spectrum\\
8&www.washingtonpost.com &11&-0.41&The Washington Post\\
9&news.yahoo.com &6&-0.394& news site provided by Yahoo!\\
10&www.alternet.org &20&-0.389 & progressive/liberal activist news service\\
11&www.dailykos.com &43&-0.384 & American political blog, liberal or progressive point of view\\
12&www.npr.org &37&-0.373&news site of the National Public Radio, a media organization that serves as a national syndicator to most public radio stations in the United States\\
13&www.cnn.com &7&-0.373&CNN\\
14&www.reuters.com &9&-0.342& Reuters\\
15&www.huffingtonpost.com &4&-0.338&liberal/progressive American news website and aggregated blog\\	
\botrule
\end{tabular}}
\end{table}

\begin{table}[!hh]
\tbl{Top 15 websites (out of top 50 most popular) with the highest $\langle e \rangle_{website}$.\label{tab:pos}}
{\begin{tabular}{@{}clccp{5cm}@{}}
\toprule
Rank & Website & Counts rank & $\langle e \rangle_{website}$ & Description/category\\
\colrule
1&blog.propertynice.com &49&0.338&properties\\
2&www.squidoo.com &15&0.310&publishing platform for posting overview material on topic of interest, e.g. "50 Things you can Reuse "\\
3&hubpages.com &34&0.189&publishing tool, all sorts of topics\\
4&www.engadget.com &41&0.186& gadgets, technology\\
5&www.collegehumor.com &28&0.138&humour\\
6&i.gizmodo.com &17&0.132& gadgets, technology\\
7&www.theonion.com &50&0.118& web site of a parody newspaper\\
8&www.flickr.com &12&0.091&image hosting and video hosting website\\
9&www.pcworld.com &39&0.078&website of a computer magazine\\
10&www.cracked.com &38&0.074&American comedy website\\
11&www.youtube.com &1&0.067&video sharing website\\
12&arstechnica.com &14&0.064& news and reviews, analysis of technology trends and expert advice\\
13&news.cnet.com &23&0.064&top technology news headlines \\
14&www.break.com &35&0.026&humor website, formerly Big-boys.com\\
15&blog.wired.com &25&0.009&aggregated blogs of Wired.com - an online technology news website\\
\botrule
\end{tabular}}
\end{table}

\begin{figure}[!ht]
\centerline{\epsfig{file=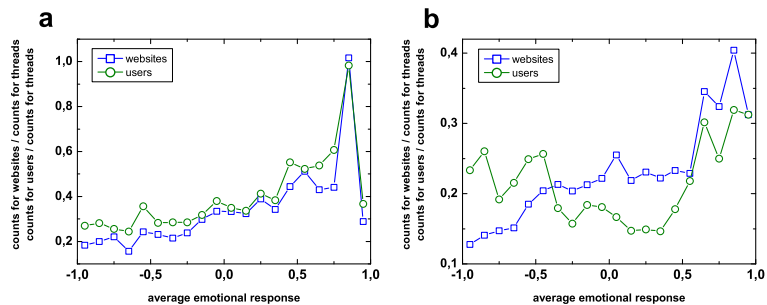,width=0.8\textwidth}}
\caption{Proportion of the average emotional response histogram of users to the histogram of threads (circles) and websites to threads (squares). (a) All data, $N_{bins}=20$. (b) Threads with a threshold $N_{comments} \geq 8$, $N_{bins}=20$.}
\label{fig:avr_histo_prop}
\end{figure}

In addition, a graph presenting average number of discussions (threads) initiated by websites grouped according to the value of average emotional response to materials coming from these websites was plotted (squares in Fig. \ref{fig:avr_diss}). It shows that most successful in terms of the number of appearance and discussions' initiation were websites that received $\langle e \rangle_{website}$ of slightly and mildly negative values. Similarly to distributions for threads and users, $\langle e \rangle_{website}$ histograms  without a threshold show 3 large peaks for values -1,0 and 1, though their proportional heights differ between the groups of threads, users and websites. After introducing a threshold a Gaussian-like distribution appears, lightly shifted to the left.

\subsection{Background removed}
In order to make more visible the differences between distributions for threads, users and websites, a procedure of background removal was carried out. By background the distribution of threads is meant here. We divide counts from $\langle e \rangle$ histogram for users/website (dark-colored bars in Fig. \ref{fig:avr_bg}a-b) by corresponding counts from $\langle e \rangle$ histogram for threads (dashed bars). The proportion is plotted for data with (Fig. \ref{fig:avr_histo_prop}a) and without a threshold in the number of comments (Fig. \ref{fig:avr_histo_prop}b). 

For data without any cut-off we get similar plots both for websites and users (squares and circles in Fig. \ref{fig:avr_histo_prop} respectively). However, there is not much point in inspecting them in much detail on their own, because as we have already seen numerous short threads tend to blur the picture. More informative are plots for  data with a cut-off. And we can see that when a threshold is introduced ($N_{comments} \geq  8$), the character of plot for websites does not change (squares in Fig. \ref{fig:avr_histo_prop}b), while plot for users transforms into one resembling U-shape (circles in Fig. \ref{fig:avr_histo_prop}b).

\begin{figure}[!ht]
\centerline{\epsfig{file=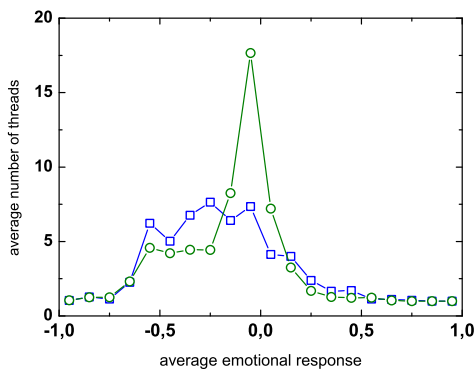,width=0.5\textwidth}}
\caption{Average number of discussions (threads) initiated by websites (squares) and users (circles) that have received a specific average emotional response. All threads with a threshold $N_{comments} \geq 8$.}
\label{fig:avr_diss}
\end{figure}

Based on the figures a number of hypotheses could be formulated. However, in order to avoid false conclusions, additional graphs have been plotted. They shed some more light on the matter under investigation. Namely, Fig. \ref{fig:avr_diss} presents average number of discussions (threads) initiated by websites (squares) and users (circles) grouped according to the value of average emotional response to materials coming from these websites/submitted by these users.

Now, let us turn back to the analysis of the results obtained in Fig. \ref{fig:avr_histo_prop}. Most interesting is the difference in graphs for websites and users, with introduced threshold. It is shown that the group of websites receiving exclusively very positive responses and groups of user receiving exclusively either very positive or very negative responses are relatively more numerous. On the other hand, the group of websites receiving exclusively very negative responses is relatively less numerous.

We could hypothesize that largely negative threads group into a relatively small number of websites which in turn originate many discussions. However, this logics fails when confronted with Fig. \ref{fig:avr_diss}. Here for values between -1 and -0.7 the average number of threads is equal or very close to 1 meaning that almost all websites which receive $\langle e \rangle$ of values from this region initiated only one thread. The same is true for the region of positive $\langle e \rangle$ (from 0.5 upwards) and for users (negative and positive regions). Also, in the case of users, those whose submitted  material received $\langle e \rangle$ of values between -0.4 and 0.5 are relatively slightly less numerous.

It is worth to emphasise that observed relations are not a special case of the threshold used ($N_{comments} \geq  8$). Starting with threads of the length of 5 or more comments, the U-shape for users clearly appears and becomes more and more evident with the increase of the threshold. Plots for websites also express the same behaviour irrespective of the threshold used.

\section{Correlations}

The objective of the following part of analysis was to determine whether emotional content obtained by computer program (both continuous and binary classifications used) correlates with the number of users' approvals (diggs up), disapprovals (diggs down) or some function thereof. Data was divided into a number of subsets for which correlation coefficients were calculated. In addition, some graphs presenting behaviour for selected ranges were plotted.

\subsection{Correlations of diggs and emotional response}

\begin{figure}[!ht]
\centerline{\epsfig{file=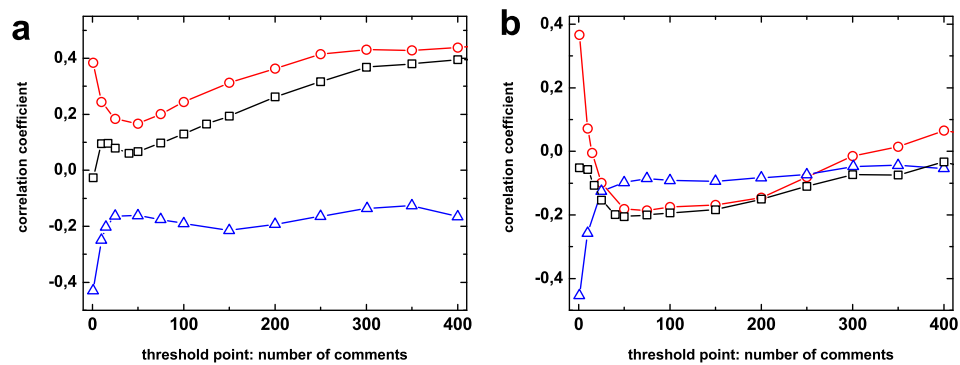,width=\textwidth}}
\caption{(a) Correlation coefficients between number of diggs $N_{diggs}$ and the average emotional response $\langle e \rangle_{thread}$ versus the threshold on the number of comments $N_{comments}$. Squares --- correlation coefficients for all threads, triangles --- correlation coefficient for threads with $\langle e \rangle_{thread} > 0$, circles ---  correlation coefficient for threads with $\langle e \rangle_{thread} < 0$. (b) Correlation coefficients between number of comments $N_{comments}$ and the average emotional response $\langle e \rangle_{thread}$ versus the threshold on the number of comments $N_{comments}$. Squares ---  correlation coefficients for all threads, (triangles) correlation coefficient for threads with $\langle e \rangle_{thread} > 0$, circles --- correlation coefficient for threads with $\langle e \rangle_{thread} < 0$.}
\label{fig:corr}
\end{figure}

For main posts (stories) only the number of approvals (diggs up) is registered by Digg.com mechanism, hence it was the only metric that could be used for determining correlation with average emotional response $\langle e \rangle_{thread}$. Correlation coefficient between number of diggs $N_{diggs}$ and $\langle e \rangle_{thread}$ for all commented stories (threads) equals to -0.027, implying no correlation on a global level. In order to establish whether or not any correlation can be observed for limited ranges of data, correlation coefficient were calculated for various data subsets obtained by introduction of different thresholds in the number of comments ('high-pass' mechanism, i.e. threads with a specific and higher number of comments were taken into account). In Fig. \ref{fig:corr}a obtained coefficients were plotted versus threshold values. For example, threshold point equal to 200 means that the coefficient was calculated for the subset of threads of the length equal to 200 comments and more.

As Fig. \ref{fig:corr}a reveals, the correlation level between $\langle e \rangle_{thread}$ and $N_{diggs}$ (squares) is positive and, except for initial bump, increases with the length of the threads (linearly for a long range of values). This behaviour is accounted for mainly by threads with negative $\langle e \rangle_{thread}$ (circles in Fig. \ref{fig:corr}a) owing to their prevailing number. Unintuitive is the fact that correlation for threads with positive $\langle e \rangle_{thread}$ is negative for all thresholds (triangles in Fig. \ref{fig:corr}a). Initially it sharply increases and from around the point equal to 25 stays at more less the same level. One would expect that more popular stories (i.e. those with larger number of diggs) will trigger more positive responses. A contrary behaviour is implied by data - popular stories tend to have lower (though, still positive) $\langle e \rangle_{thread}$. This is most significant in the region where small threads dominate - the most negative value of correlation coefficient can be observed there.

Similarly, Fig. \ref{fig:corr}b also implies that it is the stories with the value of $\langle e \rangle_{thread}$ nearing zero (either positive or negative) that attract larger numbers of diggs. Positive correlation for negative $\langle e \rangle_{thread}$ means that the higher (closer to zero) the negative values, the more diggs they obtain. In addition, correlation for two "low-pass" thresholds were also calculated and listed in Table \ref{tab:corr} proving explicitly that it is the impact of shorter threads that hides in the overall data the correlations for longer threads.

\begin{table}[ph]
\tbl{Correlation coefficients $c$ between average emotional response $\langle e \rangle_{thread}$ and the number of diggs $N_{diggs}$ versus the number of comments with an imposed threshold $N_{comments} \le x$; $M$ stands for number of threads.\label{tab:corr}}
{\begin{tabular}{@{}p{2cm}p{2cm}p{2cm}@{}}
\toprule
x & c & M \\
\colrule
100 & -0.017 & 125399\\
150 & -0.019 & 127038\\
\botrule
\end{tabular}}
\end{table}

\subsection{Correlations of comments and average emotional response}
Similar calculations to those for diggs were carried out for correlation between $\langle e \rangle_{thread}$ and the number of comments $N_{comments}$ (Fig. \ref{fig:corr}b). Correlation coefficient between the average emotional response and the number of comments is slightly negative with a minimum of 0.2 for the threshold point equal to around 50. The cut through 0 around point 500 should be treated with care as fluctuations are more probable around this region due to lower statistics. The plot implies that longer threads tend to be (slightly) more negative. In order to further investigate the relation between $\langle e \rangle_{thread}$ and $N_{comments}$, coefficients for threads with exclusively negative and positive $\langle e \rangle_{thread}$ were calculated and plotted in Fig. \ref{fig:corr}b.

Relatively high (in the case of $\langle e \rangle_{thread} < 0$) and low (in the case of $\langle e \rangle_{thread} > 0$) initial values stem from the fact that in the region where short (but solely negative/positive) comments dominate their large (in absolute values) $\langle e \rangle_{thread}$ have a huge impact, resulting in, respectively, very high (the longer the negative thread the less negative it gets) and very low (the longer the positive thread the less positive it gets) values of coefficients. As was the case with diggs, surprisingly the correlation for $\langle e \rangle_{thread} > 0$ is (slightly) negative for larger values of comment thresholds as well. For $\langle e \rangle_{thread} < 0$ an expected behaviour for the threshold range between 10 and 300 occurs - the negative value of correlation indicates that (for this range) longer threads indeed tend to be more negative. However, for exclusively long threads this tendency is not the case.  

\begin{figure}[!ht]
\centerline{\epsfig{file=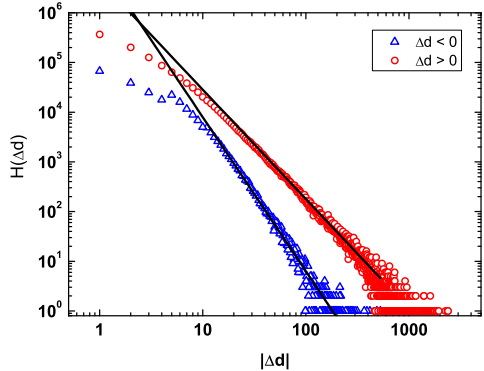,width=0.5\textwidth}}
\caption{Log-log histogram of the absolute value of digg difference $H(|\Delta d|)$. Triangles represent the branch $\Delta d < 0$ while circles $\Delta d > 0$. The solid lines represent power-law fitting to the data: in case of positive $\Delta d$ the line follows $H(|\Delta d|) \sim |\Delta d|^{-2.2}$, as for negative $\Delta d$ it is $H(|\Delta d|) \sim |\Delta d|^{-3.1}$.}
\label{fig:hdiff}
\end{figure}

In case of comments, the data consisted also the information about the number of disapprovals (diggs down, negative diggs) submitted by users. This fact allows to consider another quantity - digg difference, defined as
\begin{equation}
\Delta d = d_{up}-d_{down},
\end{equation} 
where $d_{up}$ ($d_{down}$) is, respectively, the number of diggs up (down) submitted to the comment. The histogram $H(\Delta d)$ is shown in Fig. \ref{fig:hdiff}, suggesting similar law behind the process of issuing both positive and negative diggs. However, as the largest $|\Delta d|$ for the positive branch is about 10 times the value for the negative one and taking into account that $H(|\Delta d|)$ for $\Delta d < 0$ drops down much faster than for $\Delta d > 0$ it seems that the users are much more reluctant to submit negative diggs.   

\begin{figure}[!ht]
\centerline{\epsfig{file=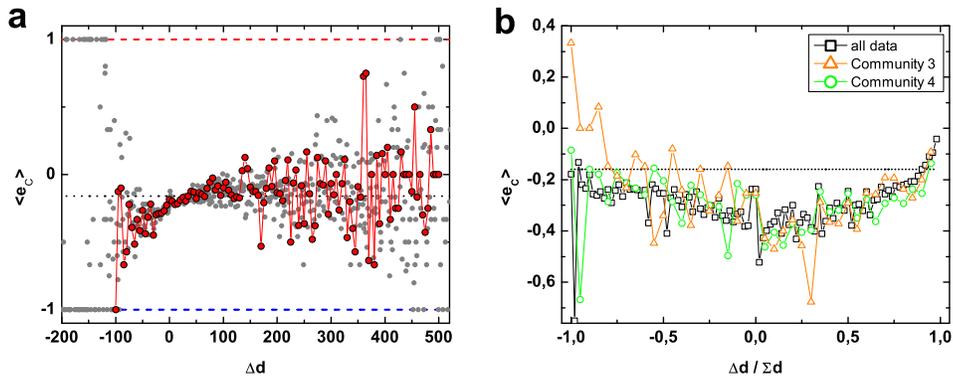,width=\textwidth}}
\caption{(a) Average emotional value of comments $\langle e_c \rangle$ and the digg difference $\Delta d$.  Grey points are real data while circles are obtained using a 5-point binning. Dotted line marks average emotional value of the whole data set $\langle e \rangle_{Digg}=-0.16$ and solid lines indicate levels $\langle e \rangle=1$ and $\langle e \rangle=-1$. (b) Average emotional value of comments $\langle e_c \rangle$ versus normalised digg difference $\frac{\Delta d}{\Sigma d}$ for all data (squares) and selected user communities (circles and triangles). Dotted line marks average emotional value of the whole data set $\langle e \rangle_{Digg}=-0.16$.}
\label{fig:diff}
\end{figure}

On this level of analysis it is also possible to check the relation between the average emotional value of comments $\langle e_c \rangle$ and the digg difference $\Delta d$. All comments that acquired a specific value of digg difference $\Delta d$ were grouped together and their average emotional value was calculated (see grey points in Fig. \ref{fig:diff}a). Figure \ref{fig:diff}a suggests that sentiment of the comments characterised with a negative value of $\Delta d$ tends to be more negative than in case of the comments with $\Delta d > 0$. Moreover, the value of $\langle e_c \rangle$ saturates for higher $\Delta d$, being close to average emotional value of the whole data set ($\langle e_c \rangle_{Digg}=-0.16$, marked with a dotted line in Fig. \ref{fig:diff}a).

One has to take into account the fact, that the values shown in Fig. \ref{fig:diff}a are not normalized. In order to present a more accurate quantity one should divide $\Delta d$ by $\Sigma d = d_{up}+d_{down}$, thus obtaining the relative value of digg difference. The results, marked with squares, are plotted in Fig. \ref{fig:diff}b, demonstrating a clear minimum around $\frac{\Delta d}{\Sigma d}=0$ and increasing values for both positive and negative $\frac{\Delta d}{\Sigma d}$ once again stopping in the vicinity of $\langle e \rangle_{Digg}=-0.16$. It leads to a rather stunning conclusion that no matter if all diggs submitted to a comment are positive or negative, its content, on average, will have a similar emotional value. A possible explanation might be put in the following way: a comment with the emotional content close to average value does not provoke a separation of opinions - it is either commonly liked or disliked. On the other hand, those comments that seem to divide the users into almost equal fractions seem to have very low $\langle e_c \rangle$. Bearing in mind that the above conclusion might be an artifact we checked the relation $\langle e_c \rangle \left( \frac{\Delta d}{\Sigma d}\right)$ for four different user communities obtained via eigenvalue spectral analysis of the weighted bipartite (i.e., users and comments) network of the most popular comments (for method details see \cite{bosa1,bosa2,bosa3}). The results, shown for two largest communities with number of comments $N_c=10214$ (circles in \ref{fig:diff}b) and $N_c=51166$ (triangles in \ref{fig:diff}b) confirm the previous observations. Although there are small discrepancies, the tendency stays the same, suggesting that the described behaviour is common regardless of the data set partition scheme.

\subsection{Correlation of comments and diggs}
\begin{figure}[!ht]
\centerline{\epsfig{file=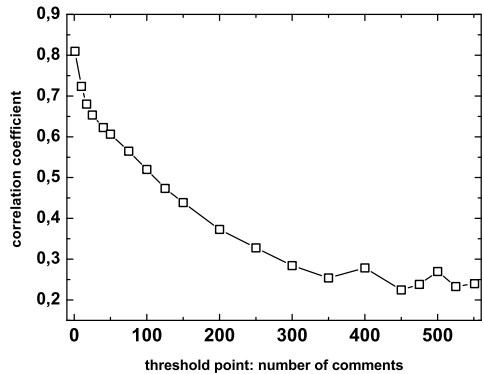,width=0.5\textwidth}}
\caption{Correlation coefficients between number of diggs $N_{diggs}$ and the number of comments $N_{comments}$ versus the threshold on the number of comments $N_{comments}$.}
\label{fig:corr_dc}
\end{figure}

The correlation between the number of diggs $N_{diggs}$ and the number of comments $N_{comments}$ (Fig. \ref{fig:corr_dc}) shows a very interesting behaviour. The coefficient is very high (over 0.8) for the data where small threads dominate and consistently decreases for longer threads, though having positive value all the time (slightly over 0.2 at the lowest). The behaviour has a very convincing heuristic explanation. Longer threads are usually developed thanks to a multiple comment exchanges between a limited number of users (usually a few, but binary exchanges - just between two people - are also frequent). Those users post additional comments, but do not digg the story again - it is not allowed by the system, even if they had such an (unlikely) wish. Hence the discrepancy in the number of diggs and comments for long discussions. For short and medium-sized threads such a duality does not occur and the numbers are roughly the same.

\section{Average response to a story}

\begin{figure}[!ht]
\centerline{\epsfig{file=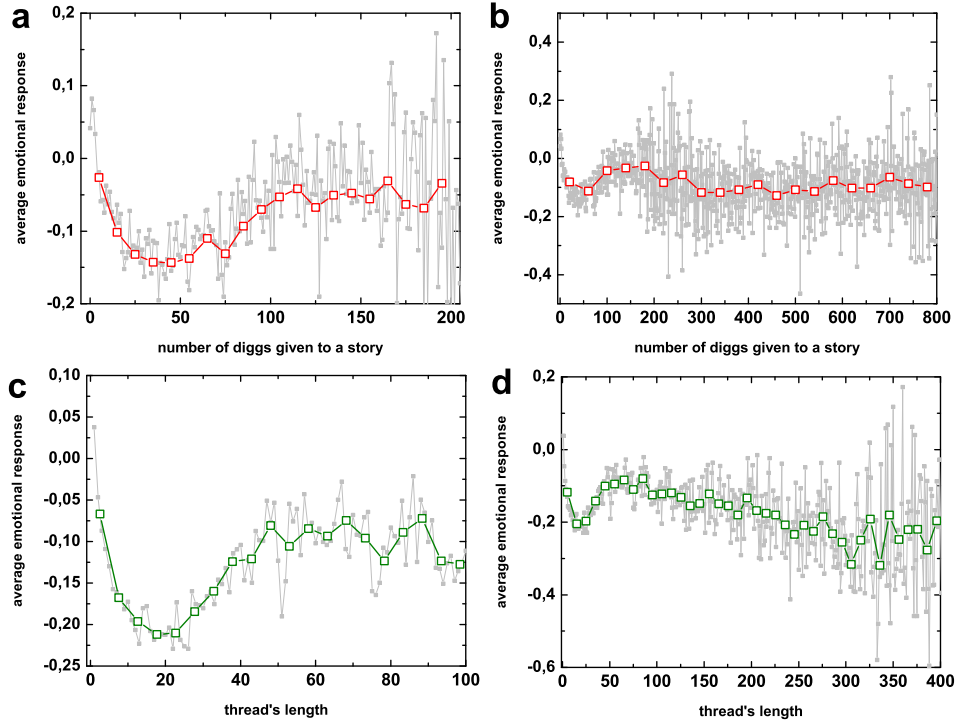,width=\textwidth}}
\caption{Average emotional response to a story $\langle e \rangle_{thread}$. (a-b) versus the number of diggs given to a story $N_{diggs}$ (c-d) versus the thread's length. Small squares are original data, big squares represent averaging with 5 bins. Plots (a,c) are close-ups of plots (b,d), respectively.}
\label{fig:aer}
\end{figure}

As a development of the analysis of diggs correlations from the previous section we examine here the dependence of average emotional response to a story $\langle e \rangle_{thread}$ on the number of diggs $N_{diggs}$ and comments $N_{comments}$ the story receives. The graphs presented in Fig. \ref{fig:aer} exhibit a very interesting behaviour. They imply that there is a specific value at which average emotional response assumes a minimum. The point in question is equal to approximately $N_{diggs}=50$ in the case of diggs (Fig. \ref{fig:aer}a) and approximately $N_{comments}=20$ in the case of comments (Fig. \ref{fig:aer}c).

In the further part the graphs differ: $\langle e \rangle_{thread}(N_{diggs})$ fluctuates around a fixed value (Fig. \ref{fig:aer}b) whereas $\langle e \rangle_{thread}(N_{comments})$ visibly decreases towards the end (Fig. \ref{fig:aer}d). The behaviour for comments is of great importance as it indicates that (beyond certain length) longer threads tend to be more negatively charged. Hence we could assume that it is the negative comments that fuel the communication. Similar conclusions were reached in \cite{sobkowicz} where it was observed that confrontational or abusive discussions lasted much longer than neutral ones. It is worth to mention that the overall behaviour of the two graphs comply with the previous results concerning correlation between the number of diggs and the number of comments. As was presented in Fig. \ref{fig:corr_dc} for short threads the correlation is very high and substantially decreases for longer discussions. That is also the case here: the assumption of minimal value for short threads is observed in plots for diggs and comments alike while the behaviour of the graphs towards longer discussions does not match.

\begin{figure}
\centerline{\epsfig{file=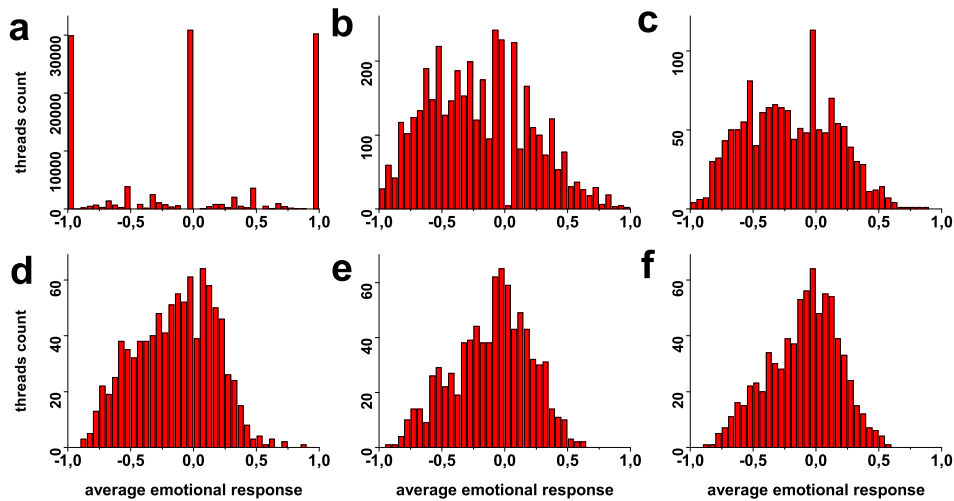,width=\textwidth}}
\caption{Average emotional response histograms for threads of specified length $N$: (a) $N \in [0,10]$, (b) $N \in [11,20]$, (c) $N \in [21,20]$, (d) $N \in [31,40]$, (e) $N \in [41,50]$, (f) $N \in [51,60]$.}
\label{fig:aer_hist}
\end{figure}

In order to explain the phenomena of minimal value seen for shorter threads, average response histograms for the initial groups of threads were plotted in Fig. \ref{fig:aer_hist}. It can be assumed that the ratio of probability that all (or almost all) comments in a thread are negative to an analogous likelihood for a positive dominance is responsible for the behaviour in question. For very short threads probabilities for extreme values of $\langle e \rangle_{thread}$ are roughly the same (cf equal bars in Fig. \ref{fig:aer_hist}a); for threads between 10-20 a major dominance for very low negative values is observed culminating for threads of the length of around 20 comments. For still longer discussions the dominance tends to diminish as the probability of obtaining solely negative comments in longer threads is very low. Around the value of 50 comments both probabilities for extreme $\langle e \rangle_{thread}$ values are again similar (this time both are very close to zero), though the graph as such is slightly shifted to the left.

\section{Influence of emotions on thread's life and end}

The issue under examination was how emotional content of comments changed with the development of a thread and whether or not its level at the beginning of a thread had any impact on how long the thread lived.

\begin{figure}[!ht]
\centerline{\epsfig{file=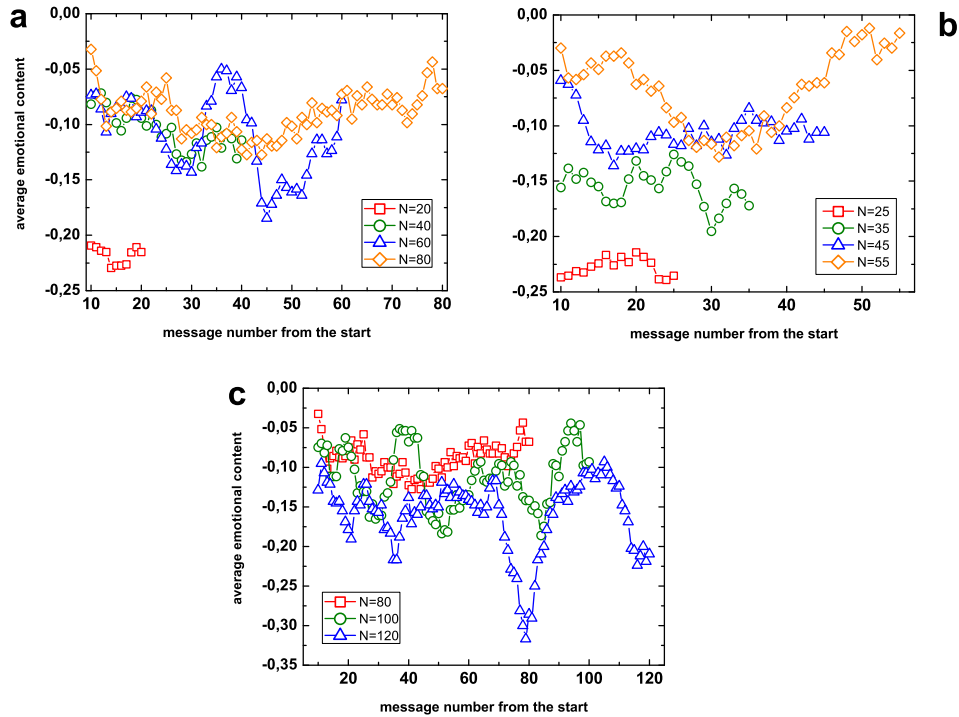,width=\textwidth}}
\caption{Average emotional content in the thread $\langle e \rangle_{thread}$ versus the message number from the start. Each point is calculated as a moving average over the last 10 comments. (a) Threads of length 20 (squares), 40 (circles), 60 (triangles) and 80 (diamonds). (b) Threads of length 25 (squares), 35 (circles), 45 (triangles) and 55 (diamonds). (c) Threads of length 80 (squares), 100 (circles) and 120 (triangles).}
\label{fig:av_inth}
\end{figure}

\subsection{Comment sequence}
\begin{figure}[!ht]
\centerline{\epsfig{file=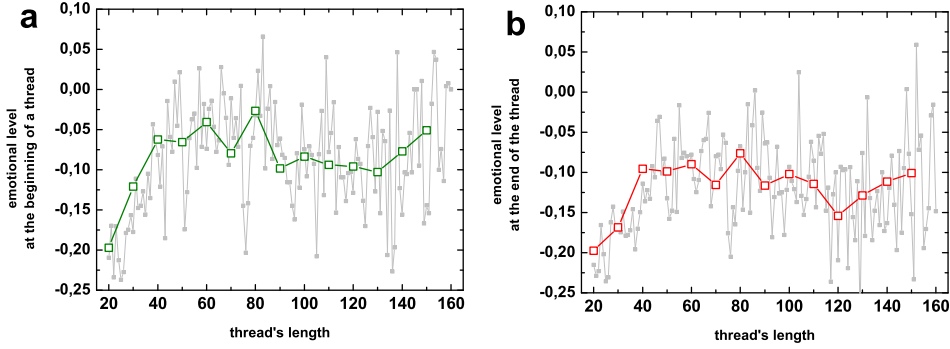,width=\textwidth}}
\caption{Emotional level versus thread's length at the beginning i.e. in the first 10 comments (a) and at the end of the thread i.e. in the last 10 comments (b). Grey points are original data, squares represent averageing with 5 bins.}
\label{fig:beg_end}
\end{figure}

The approach concentrated on the sequence of comments in a thread, regardless of how much time elapsed between publishing consecutive comments. The following procedure was applied: threads of the same size were grouped together and, starting from the 10th comment in a thread, for each point a moving average of the emotional content of the last 10 comments was calculated. Graphs for a few selected groups of threads are presented in Fig. \ref{fig:av_inth}. Figure \ref{fig:av_inth}a (and very clearly Fig. \ref{fig:av_inth}b) suggests that average emotional content both at the beginning and at the end of a thread increases with the increase of thread's length. The series start with negative values and grow towards zero with the thread length. However, Fig. \ref{fig:av_inth}c implies that this may be the case only for shorter threads; graphs for threads of the length of 80, 100 and 120 comments do not show the regularity. Figure \ref{fig:beg_end} presenting the emotional level versus thread's length at the beginning (a) and at the end (b) of a thread supports the observation. The increase is present only for threads of the length between 20 and 60 comments, then we observe a saturation. One feasible heuristic explanation for this behaviour is that when the initial comments are emotionally more negative the thread relatively quickly wears out, because the participants give vent to their emotions early on in the thread and later do not have enough emotional potential to carry on discussion.

Bearing in mind previous results one can also notice that the behaviour of emotional level at the beginning and at the end of a thread for the analysed in Fig. \ref{fig:beg_end} range of thread's lengths (20-160 comments) resembles to large extent the behaviour of average emotional response to stories for that very range (cf Fig. \ref{fig:aer}): an initial increase for the range between 20 and 60 is followed by saturation (or very gradual decrease in the case of average emotional response) for values between 60 and 160. This may imply that emotional content of the first 10 and the last 10 comments in a thread is to some degree representative of the overall emotional content of a thread, at least for the range of thread's lengths in question.

\begin{figure}[!ht]
\centerline{\epsfig{file=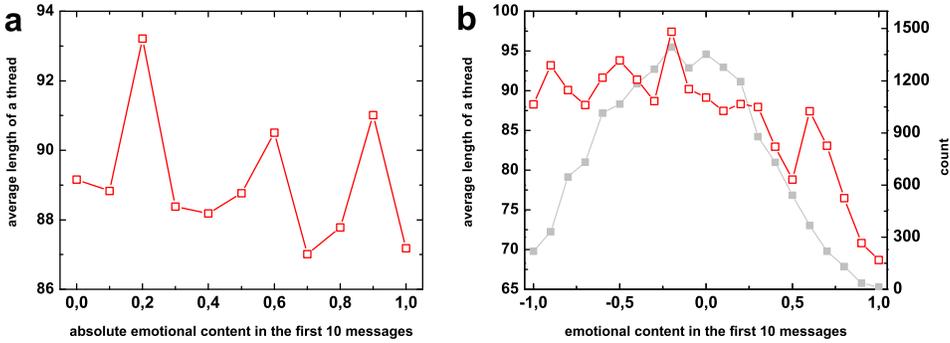,width=\textwidth}}
\caption{Average length of a thread $\langle L \rangle$ versus the absolute emotional value in the first 10 comments $| \langle e \rangle_{10} |$ (a), the emotional value in the first 10 comments $\langle e \rangle_{10}$. Grey points in plot (b) represent number threads with a specific emotional value.}
\label{fig:th_length}
\end{figure}

In order to more fully determine whether emotional content of the first 10 comments influences the length of the thread, averages of thread's length for different levels of values of initial emotions were calculated. For this analysis all threads having 10 or more comments were taken into account. 

Initially, an approach using the absolute value of initial emotions was applied. As Fig. \ref{fig:th_length}a shows, in such a case values of average thread's length vary little and one can't say that highly emotional launch (be it negative or positive) generally leads to a longer or shorter discussion. However, taking into account the whole range of possible values (between -1 and 1) reveals more information (Fig. \ref{fig:th_length}b).

One should start analysing Fig. \ref{fig:th_length}b with an observation that, in addition to average thread's length, plotted in grey are the counts of instances that fall into particular bins. For the two most positive values of average emotional content these counts are small and hence the last two points should be treated with care. Nonetheless, even when the last two points are not taken into account an interesting trend can be noticed. Namely, for negative and slightly positive initial emotional content (up to the value of 0.3) one observes a stable level of average thread's length, then for values between 0.3 and 0.8 a drop is clearly visible. This implies that in the population of all threads of the length of 10 comments and more it is the mildly and highly positive launches that lead to shorter discussions. The intuitive explanation based on everyday observation is that when people agree there is not really much point in carrying on the conversation.

The result is not, however, in agreement with the one obtained earlier for the population of threads of the length between 20 and 60 comments (Fig. \ref{fig:aer}; smaller values of average initial emotional content plotted for shorter threads). This implies that while the regularity described in the previous paragraph holds true for the whole population, it may not apply to particular subsets and local patterns (even reverse) may occur.

\section{Conclusions}

During the period of analysis a number of possible study directions were penetrated. Some provided interesting results. Below are listed conclusions deemed by the authors as bearing greatest significance.

It was established that once a certain length of the thread is reached, the regularity that longer threads acquire more negative charge is valid. We can assume that it is the negative emotions that (starting form some point) propel the discussion in longer threads. However, for the thread to develop, for certain lengths it should not be launched with highly negative emotions. If the first comments are largely negative the thread dies quickly. A possible explanation of this mechanism is that when participants give vent to their emotions early on in the thread, they later do not have enough emotional potential to carry on discussion. Similarly, mildly and highly positive launches tend to lead to shorter discussions. This suggests presence of two different mechanisms governing the evolution of the discussion and, consequently, its length.

With the use of averages it was ascertained that the most negative emotional responses were prompted by websites dealing with politics. On the other hand, those concerned with technology, giving advice or humour generated most positive reactions. The most successful in terms of the number of appearance and discussions' initiation were websites that received average response of slightly and mildly negative values.

Contrary to expectations, no correlation was found between the number of diggs received by a comment and its emotional charge. This leads to a conclusion that \textit{digging} and \textit{burying} are driven by the interest of the comment, rather than by its emotional content. People vote for comments that are interesting, witty or say something familiar and disapprove of those which are boring, irrespective of the emotions they convey.  

While analysing correlations on a story level, an interesting behaviour of the relation between the number of diggs and the number of comments received by a story was found. The correlation between the two quantities is high for data where small threads dominate and consistently decreases for longer threads (thought staying positive all the time). This behaviour has a convincing explanation. Namely, longer threads are formed as a result of exchanges between a limited number of discussion participant. No matter how many comments these users write in a specific thread, they digg the story only once. Hence the longer the thread, the wider the discrepancy between the number of diggs and the number of comments.

Results indicate that threads with small number of diggs (corresponding to small number of comments) are relatively  more objective. With the increase in the length and the number of diggs the threads' subjectivity increases.

It is worth to highlight that a considerable number of distributions plotted for Digg.com followed (at least for some ranges of values) power laws. This included comments and diggs distributions as well as users' productivity. Thus it was confirmed also is this work that scale-free relations are often found in data concerning online behaviour.

\section*{Acknowledgments}
The work was supported by EU FP7 ICT Project Collective Emotions in Cyberspace CYBEREMOTIONS and Polish Ministry of Science Grant 1029/7.PR UE/2009/7. J.S. and J.A.H acknowledge the support from the European COST Action MP0801 Physics of Competition and Conflicts and from the Polish Ministry of Science Grant No. 578/NCOST/2009/0. J.S. acknowledges support from a special grant of the Dean of the Faculty of Physics, WUT.

\bibliographystyle{ws-acs}
\bibliography{ws-acs}

\begin{thebibliography}{10}
\bibitem{ball} Ball P., {\it Critical Mass: How One Thing Leads to Another}. (Farrar, Straus and Giroux; London, 2004).
\bibitem{cast} Castellano C., Fortunato S., Loreto V, Statistical physics of social dynamics. \textit{Rev. Mod. Phys.} {\bf 81} (2009) 591.
\bibitem{axel} Axelrod R., The Dissemination of Culture: A Model with Local Convergence and Global Polarization. \textit{J. Conflict Resolut} {\bf 41} (1997) 203.
\bibitem{sznajd} Sznajd-Weron K., Sznajd J., Opinion Evolution in Closed Community. \textit{Int. J. Mod. Phys. C} {\bf 11} (2000) 1157-1165.
\bibitem{bosa1} Mitrovi\'{c} M., Tadi\'{c} B., Bloggers behavior and emergent communities in blog space. \textit{Eur. Phys. J. B} {\bf 73} (2010) 293-301.
\bibitem{frank} Schweitzer F., Garcia D., An Agent-Based Model of Collective Emotions in Online Communities. \textit{Eur. Phys. J. B} {\bf 77}, (2010) 533–545.
\bibitem{bosa2} Mitrovi\'{c} M., Paltoglou G., Tadi\'{c} B., Networks and emotion-driven user communities at popular blogs. \textit{Eur. Phys. J. B} {\bf 77} (2010) 597–609.
\bibitem{bosa3} Mitrovi\'{c} M., Paltoglou G., Tadi\'{c} B., Quantitative analysis of bloggers’ collective behavior powered by emotions. {\it J. Stat. Mech.} (2011) P02005.
\bibitem{ania2} Chmiel A., Sobkowicz P., Sienkiewicz J., Paltoglou G., Buckley K., Thelwall M., Ho{\l}yst J. A., Negative emotions boost user activity at BBC forum, {\it Physica A} {\bf 390} (2011) 2936–2944.
\bibitem{ania3} Chmiel A., Sienkiewicz J., Paltoglou G., Buckley K., Thelwall M., Kappas A., Ho{\l}yst J. A., Collective Emotions Online and Their Influence on
Community Life, {\it PLoS ONE} {\bf 6(7)} (2011) e0022207.
\bibitem{weronski} Wero\'{n}ski P., Sienkiewicz J., Paltoglou G., Buckley K., Thelwall K., Ho{\l}yst J. A., Emotional Analysis of Blogs and Forums Data. \textit{e-print arXiv} (2011) 1108.5974.
\bibitem{frank2} Garcia D., Garas A., Schweitzer F., Positive words carry less information than negative words. \textit{e-print arXiv} (2011) 1110.4123.
\bibitem{aga} Czaplicka A., Chmiel A., Ho{\l}yst J. A., Emotional Agents at the Square Lattice. \textit{Acta Phys. Pol. A} {\bf 117(4)} (2010) 688-694.
\bibitem{ania1} Chmiel A., Ho{\l}yst J.A. Flow of emotional messages in artificial social networks. {\it Int. J. Mod. Phys. C} {\bf 21} (2010) 593–602.
\bibitem{rank} Rank S., Docking Agent-based Simulation of Collective Emotion to Equation-based Models and Interactive Agents. \textit{Proceedings of Agent-Directed Simulation Symposium, 2010 Spring Simulation Conference.} (2010) 82-89.
\bibitem{app1} Gobron S., Ahn J., Paltoglou G., Thelwall M., Thalmann D., {\it Vis. Comput.} {\bf 26} (2010) 505.
\bibitem{app2} Skowron M., Pirker H., Rank S., Paltoglou G., Gobron S., in {\it Proceedings of the 24th International FLAIRS Conference}, (AIII Press 2011).
\bibitem{app3} Skowron M., Rank S., Theunis M., Sienkiewicz J., {\it LNCS} {\bf 6974} (2011) 337.
\bibitem{mike1} Thelwall M., Wilkinson D., Uppal S., Data mining emotion in social network communication: Gender differences in MySpace. In \textit{Journal of the American Society for Information Science and Technology} {\bf 61} (2010) 190-199.  
\bibitem{sobkowicz} Sobkowicz P., Sobkowicz A.. Dynamics of hate based Internet user networks. \textit{Eur. Phys. J. B}, {\bf 73} (2010) 633-643.
\bibitem{pang} Pang B., Lee L., Opinion mining and sentiment analysis. In \textit{Foundation and Trends in Information Retrieval} {\bf 2} (2008) 1-135. 
\bibitem{hLM} F. Sebastiani, Machine Learning in Automated Text Categorization. {\it ACM Computing Surveys} {\bf 34} (2002) 1-47.
\bibitem{blogs} I. Ounis, C. Macdonald, I. Soboroff, in Proceedings of the Second International Conference on Weblogs and Social Media (2008).
\bibitem{lm1} F. Peng, D. Schuurmans, and S. Wang, Language and task independent text categorization with simple language models. in  {\it NAACL '03} (2003) 110-117.
\bibitem{digg} {\it http://www.digg.com}.
\bibitem{lerman1} K. Lerman and A. Galstyan, Analysis of social voting patterns on digg. In \textit{Proceedings of WOSP} (2008) 7-12.
\bibitem{zhu} Zhu Y., Measurement and analysis of an online content voting network: a case study of Digg. In \textit{Proceedings of the 19th international conference on WWW}, (2010) 1039-1048.
\bibitem{szabo} Szabo G., Huberman B., Predicting the popularity of online content, {\it Communications of the ACM} {\bf 53(8)} (2010) 80-88.
\bibitem{lerman2} Lerman K., Ghosh R., Information contagion: an empirical study of the spread on news on Digg and Twitter social networks, in {\it Proceedings of the Fourth International AAAI Conference on Weblogs and Social Media} (2010) 90-97.
\bibitem{rangwala} Rangwala H., Jamali S., Defining a coparticipation network using comments on Digg, {\it IEEE Intelligent Systems} {\bf 25(4)} (2010) 36-44.
\bibitem{krap} Krapivsky P., Redner S., Leyvraz F., Connectivity of growing random networks, {\it Phys. Rev. Lett.} {\bf 85} (2000) 4629. 
\bibitem{ba} Barab\'{a}si A.-L., Albert R., Emergence of scaling in random networks. \textit{Science} {\bf 286} (1999) 509-512.
\bibitem{kr} Krapivsky P., Redner S., Organization of growing random networks, {\it Phys. Rev. E} {\bf 63} (2001) 066123.
\end{thebibliography}

\end{document}